\documentclass[a4paper]{ring}
\usepackage{natbib}
\usepackage{graphicx}
\idline{288}{39}
\begin{document}
   \title{The Magnificent Seven: Nearby Isolated Neutron Stars with strong Magnetic Fields
}

   \author{F. Haberl
}


   \institute{Max-Planck-Institut f{\"u}r extraterrestrische Physik,
              Giessenbachstra{\ss}e, 
              85748 Garching, Germany, \email{fwh@mpe.mpg.de}
             }

   \authorrunning{F. Haberl}  
   \titlerunning{Nearby Isolated Neutron Stars with strong Magnetic Fields} 
   \maketitle
%

\section{Introduction}

Estimates from the current pulsar birth rate and the number of supernovae to
account for the heavy-element abundance suggest that in total 10$^8$ to 10$^9$
isolated neutron stars exist in our Galaxy. Only a small fraction is detectable 
as young neutron stars due to their emission of thermal X-rays (for $\sim$10$^6$
years) or pulsed radio emission (for $\sim$10$^8$ years). Proposals that an old
''recycled" population of isolated neutron stars re-heated by accretion from the 
interstellar medium (ISM) should be detectable by ROSAT, triggered several projects to search
for such objects in the ROSAT all-sky survey data. Over the last decade seven
very soft X-ray sources with particular characteristics were discovered in ROSAT data.
Extreme X-ray to optical flux ratios and low absorption column densities strongly suggest
that these objects are nearby isolated neutron stars 
\citep[see reviews by][]{2000PASP..112..297T,2001xase.conf..244M,haberl2004COSPAR}.
Using HST, parallax measurements yield a distance of 117$\pm$12 pc for RX\,J1856.5$-$3754
\citep{2002ApJ...576L.145W}.
The detection of relatively high proper motion (PM) for the three brightest stars makes 
accretion from the ISM highly ineffective and favours the picture of cooling neutron stars 
with an age of $\sim$10$^5$--10$^6$ years to power the X-rays.
Tracing back the apparent trajectories suggests that the brightest of the ROSAT-discovered isolated
neutron stars were born in the Sco OB2 complex which is the closest OB association 
\citep[see e.g.][~and references therein]{2005A&A...429..257M}.

\begin{figure*}
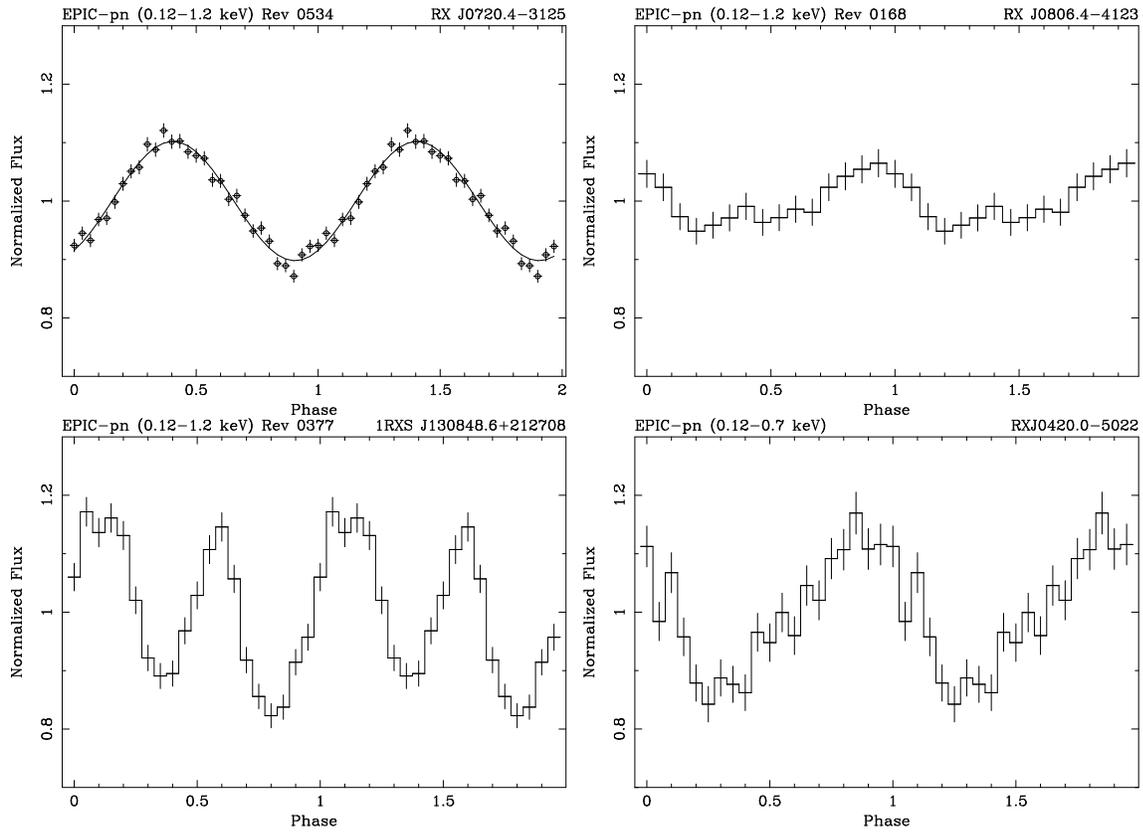

   \centering
   \resizebox{\hsize}{!}{{\includegraphics[angle=-90,clip=true]{fhaberl_pulse0720.ps}}
                          \hspace{5mm}
                          \includegraphics[angle=-90,clip=true]{fhaberl_pulse0806.ps}}
   \vspace{5mm}
   \resizebox{\hsize}{!}{{\includegraphics[angle=-90,clip=true]{fhaberl_pulse1223.ps}}
                          \hspace{5mm}
                          \includegraphics[angle=-90,clip=true]{fhaberl_pulse0420.ps}}
   \caption{EPIC-pn light curves folded with the pulse period of four thermal isolated 
            neutron stars. For direct comparison of the pulse profile the flux is normalized
	    to the mean and plotted on the same scale. Except for RX\,J0420.0$-$5022, which
	    shows the softest X-ray spectrum, data from the same energy band was used. To
	    gain statistics data of four observations of RX\,J0420.0$-$5022 were merged
	    \citep{2004A&A...424..635H}.}
   \label{fig-pulses}
\end{figure*}

The X-ray spectra of the ``magnificent seven", as they are sometimes called in the literature,
are thermal and blackbody-like without a hard power-law tail as it is often observed in other
isolated neutron stars \citep[e.g.][]{2002nsps.conf..273P}. Typical observed blackbody temperatures kT
are in the range of 40$-$110 eV (see Table 1). From five stars X-ray pulsations were detected with
pulse periods between 3 s and 12 s and pulsed fractions between 4\% and 18\% (Fig.\,\ref{fig-pulses}). 
However, for the X-ray brightest star RX\,J1856.5$-$3754 no pulsations with a stringent 
upper limit on periodic variation of 1.3\% (2$\sigma$ confidence level in the 0.02 - 1000 s range) 
were found \citep{2003A&A...399.1109B}. A surprising discovery was that the X-ray spectrum and the 
pulsed fraction observed from RX\,J0720.4$-$3125 changes on a time-scale of years which may be 
caused by precession of the neutron star \citep{2004A&A...415L..31D}.

\begin{table*}
\caption[]{X-ray and optical properties of the magnificent seven}
\begin{tabular}{lcccccc}
\hline\noalign{\smallskip}
\multicolumn{1}{l}{Object} &
\multicolumn{1}{c}{kT} &
\multicolumn{1}{c}{Period} &
\multicolumn{1}{c}{Amplitude} &
\multicolumn{1}{c}{Optical} &
\multicolumn{1}{c}{PM} &
\multicolumn{1}{c}{Ref.} \\

\multicolumn{1}{l}{} &
\multicolumn{1}{c}{eV} &
\multicolumn{1}{c}{s} &
\multicolumn{1}{c}{\%} &
\multicolumn{1}{c}{mag} &
\multicolumn{1}{c}{mas/year} &
\multicolumn{1}{c}{} \\
\noalign{\smallskip}\hline\noalign{\smallskip}
RX\,J0420.0$-$5022   &  44    &  3.45 & 13     & B = 26.6		&     & 1   \\
RX\,J0720.4$-$3125   &  85-95 &  8.39 & 8-15   & B = 26.6		&  97 & 2,3,4,5,6 \\
RX\,J0806.4$-$4123   &  96    & 11.37 &  6     & B $>$ 24		&     & 7,1 \\
RBS\,1223$^{(1)}$    &  86    & 10.31 & 18     & m$_{\rm 50ccd}$ = 28.6 &     & 8,9,10,11 \\
RX\,J1605.3+3249     &  96    & $-$   & $-$    & B = 27.2		& 145 & 12,13,14,15 \\
RX\,J1856.5$-$3754   &  60    & $-$   & $<$1.3 & V = 25.7		& 332 & 16,17,18 \\
RBS\,1774$^{(2)}$    & 101    & 9.44  &  4     & R $>$ 23		&     & 19,20   \\
\noalign{\smallskip}\hline\noalign{\smallskip}
\end{tabular}

$^{(1)}$ = 1RXS\,J130848.6+212708\\
$^{(2)}$ = 1RXS\,J214303.7+065419\\
References: 
(1) \citet{2004A&A...424..635H}
(2) \citet{1997A&A...326..662H}
(3) \citet{2001A&A...365L.302C}
(4) \citet{2004A&A...419.1077H}
(5) \citet{2004A&A...415L..31D}
(6) \citet{2003A&A...408..323M}
(7) \citet{2002A&A...391..571H}
(8) \citet{1999A&A...341L..51S}
(9) \citet{2002A&A...381...98H}
(10) \citet{2002ApJ...579L..29K}
(11) \citet{2003A&A...403L..19H}
(12) \citet{1999A&A...351..177M}
(13) \citet{2003ApJ...588L..33K}
(14) \citet{2004ApJ...608..432V}
(15) \citet{2005A&A...429..257M}
(16) \citet{1997Natur.389..358W}
(17) \citet{2002ApJ...576L.145W}
(18) \citet{2003A&A...399.1109B}
(19) \citet{2001A&A...378L...5Z}
(20) \citet{2005astroph0503239}
\end{table*}

\begin{table*}
\caption[]{Magnetic field estimates}
\begin{tabular}{lcccc}
\hline\noalign{\smallskip}
\multicolumn{1}{l}{Object} &
\multicolumn{1}{c}{dP/dt} &
\multicolumn{1}{c}{E$_{\rm cyc}$} &
\multicolumn{1}{c}{B$_{\rm db}$} &
\multicolumn{1}{c}{B$_{\rm cyc}$} \\

\multicolumn{1}{l}{} &
\multicolumn{1}{c}{10$^{-13}$ ss$^{-1}$} &
\multicolumn{1}{c}{eV} &
\multicolumn{1}{c}{10$^{13}$ G} &
\multicolumn{1}{c}{10$^{13}$ G} \\
\noalign{\smallskip}\hline\noalign{\smallskip}  
RX\,J0420.0$-$5022   &  $<$92       & 330?      & $<$18     & 6.6?      \\
RX\,J0720.4$-$3125   &  1.4$\pm$0.6 & 260       & 2.8$-$4.2 & 5.2       \\
RX\,J0806.4$-$4123   &  $<$18       &           & $<$14     &           \\
RBS\,1223            &  $<$9        & 100-300   & $<$10     & 2$-$6     \\
RX\,J1605.3+3249     &              & 450-480   &           & 9.1$-$9.7 \\
RX\,J1856.5$-$3754   &              &           & $\sim$1   & 	        \\
RBS\,1774            &              & $\sim$700 &           & $\sim$14  \\
\noalign{\smallskip}\hline\noalign{\smallskip}
\end{tabular}
\end{table*}

%

\section {Broad absorption lines}

\begin{figure*}
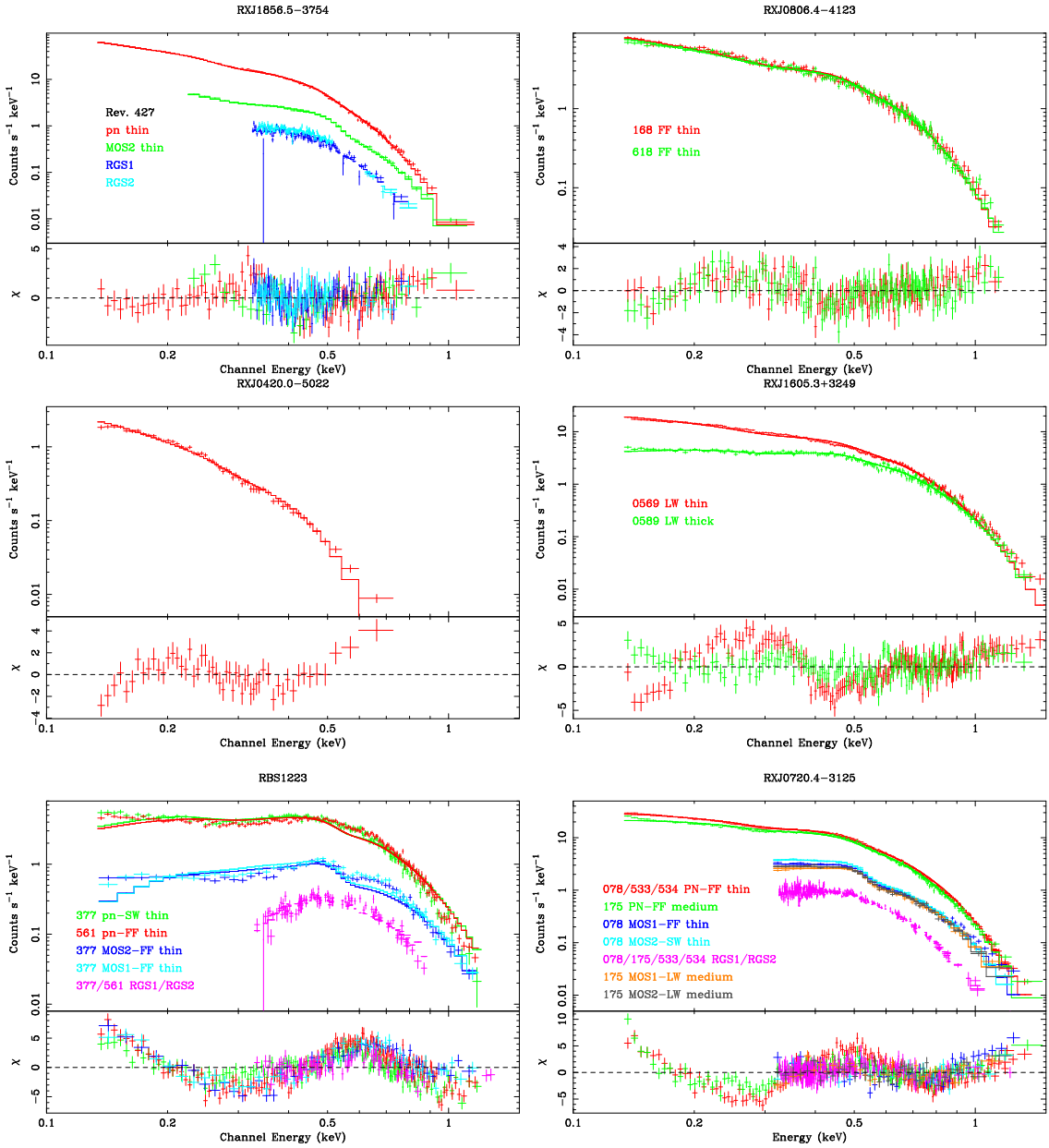

   \centering
   \resizebox{\hsize}{!}{{\includegraphics[angle=-90,clip=true]{fhaberl_spec1856.ps}}
                          \hspace{3mm}
                          \includegraphics[angle=-90,clip=true]{fhaberl_spec0806.ps}}
   \vspace{3mm}
   \resizebox{\hsize}{!}{{\includegraphics[angle=-90,clip=true]{fhaberl_spec0420.ps}}
                          \hspace{3mm}
                          \includegraphics[angle=-90,clip=true]{fhaberl_spec1605.ps}}
   \vspace{3mm}
   \resizebox{\hsize}{!}{{\includegraphics[angle=-90,clip=true]{fhaberl_spec1223.ps}}
                          \hspace{3mm}
                          \includegraphics[angle=-90,clip=true]{fhaberl_spec0720.ps}}
   \caption{EPIC-pn and RGS spectra of thermal isolated neutron stars fitted with an absorbed 
            blackbody model. The fits represent the calibration status as available with SAS release
	    6.0.0.}
        \label{fig-bbfits}
    \end{figure*}

XMM-Newton observations of the thermal isolated neutron stars revealed deviations 
from the Planckian shape in the X-ray spectra obtained by the EPIC-pn and RGS instruments.
Fig.\,\ref{fig-bbfits} shows a comparison of the EPIC-pn spectra of the six best
observed thermal isolated neutron stars fitted with an absorbed blackbody model.
Large residuals are seen from RBS\,1223 \citep{2003A&A...403L..19H}, 
RX\,J0720.4$-$3125 \citep{2004A&A...419.1077H} and RX\,J1605.3+3249, in the latter
case the deviations were discovered in RGS spectra \citep{2004ApJ...608..432V}. 
Non-magnetic neutron star atmosphere models 
\citep[e.g.][]{2002A&A...386.1001G,2002nsps.conf..263Z} can not explain the X-ray spectra:
Iron and solar mixture atmospheres
cause too many absorption features and deviations from a blackbody model in particular
at energies between 0.5 and 1.0 keV which are not seen in the measured spectra. On the other 
hand the spectrum of a pure hydrogen model is similar in shape to that of a blackbody and 
does not fit the data either. Moreover, hydrogen atmosphere models over-predict the 
actually observed optical fluxes by large factors \citep[$\sim$300, see ][]{2002nsps.conf..263Z}.

\begin{figure}
   \centering
   \resizebox{\hsize}{!}{\includegraphics[angle=-90,clip=true]{fhaberl_spec1223l.ps}}
   \vspace{3mm}
   \resizebox{\hsize}{!}{\includegraphics[angle=-90,clip=true]{fhaberl_spec0720l.ps}}
   \vspace{3mm}
   \resizebox{\hsize}{!}{\includegraphics[clip=true]{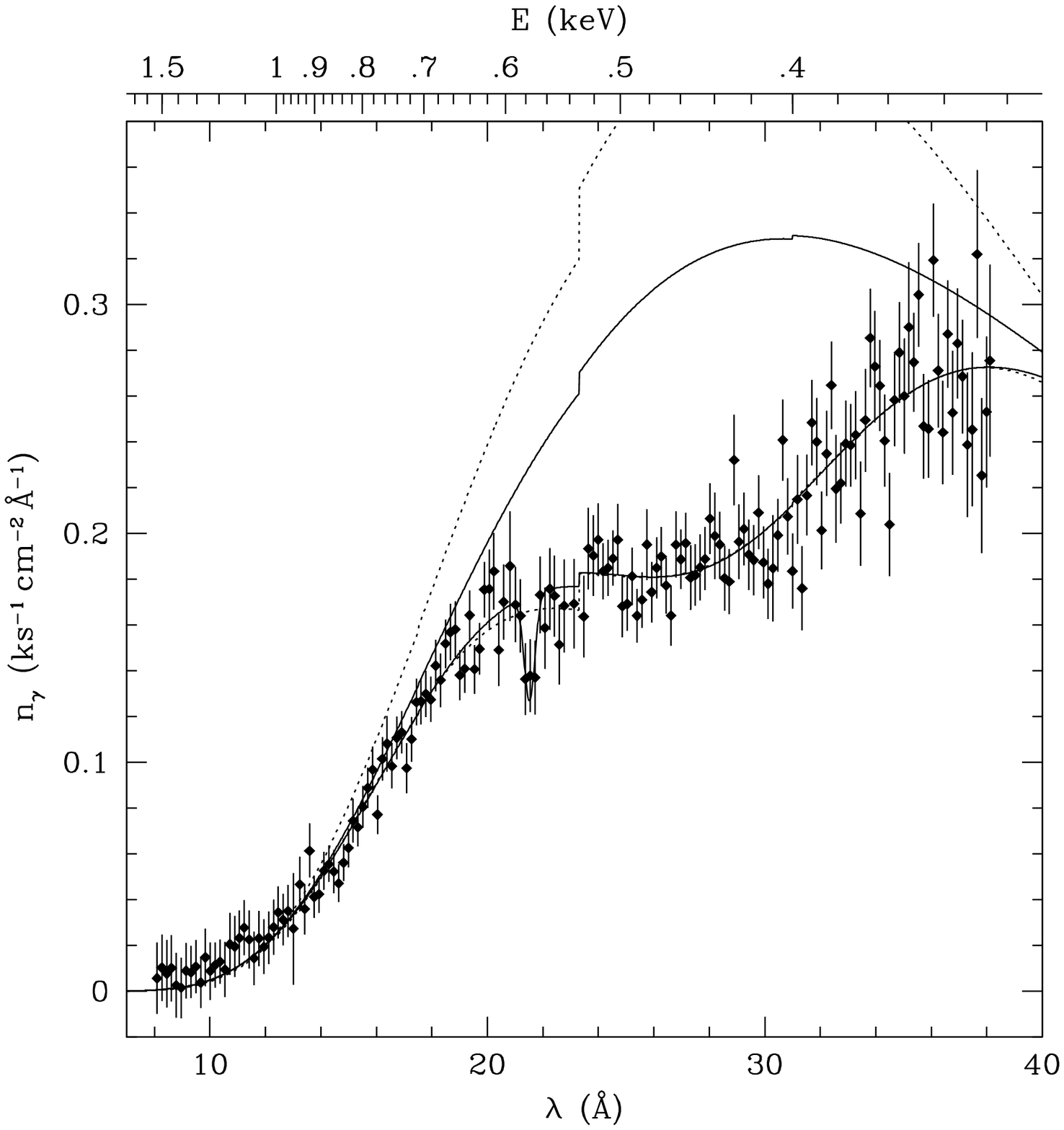}}
   \caption{Top two panels: EPIC-pn and RGS spectra of two thermal isolated neutron stars fitted 
            with an absorbed 
            blackbody model including a broad absorption line. Bottom panel: RGS spectrum of 
	    RX\,J1605.3+3249 modeled with blackbody continuum and broad absorption line. A 
	    possible additional narrow absorption line at 21.5\AA\ is also included 
	    \citep[from][]{2004ApJ...608..432V}.}
        \label{fig-gaussfits}
    \end{figure}

The XMM-Newton spectra  can best be modeled with a Planck continuum  
including a broad, Gaussian shaped absorption line 
(Fig.\,\ref{fig-gaussfits}). Line centroid energies are summarized in
Table\,2. In the EPIC-pn data of RBS\,1223 \citep[see also ][]{2005A&A...INS} 
and RX\,J0720.4$-$3125 
the depth of the absorption line (or the equivalent width) was 
found to vary with pulse phase. In the cases of RX\,J0806.4$-$4123 and RX\,J0420.0$-$5022 it
is not clear to which extent the residuals of the blackbody fits are caused by systematic 
calibration uncertainties \citep{2004A&A...424..635H}. 
In particular RX\,J0806.4$-$4123 shows a residual pattern similar to that of 
RX\,J1856.5$-$3754 which is believed to exhibit a pure blackbody spectrum as seen from the high
resolution Chandra LETGS spectrum \citep{2003A&A...399.1109B}. For RBS\,1774 recently an absorption
feature at $\sim$0.7 keV was reported from the analysis of EPIC spectra \citep{2005astroph0503239}.
At such high energies (the highest line energy reported from the thermal isolated neutron stars)
the energy resolution is better and the calibration uncertainties smaller than at lower energies.

%

%

\section{Strongly magnetized neutron stars}

An H$_{\alpha}$ emission line nebula was discovered around RX\,J1856.5$-$3754 
\citep{2001A&A...380..221V}. Assuming that magnetic dipole breaking powers this nebula
and using an age of the star of 5 x 10$^5$ years \citep{2002ApJ...576L.145W} allows 
an estimate of the magnetic field strength of the neutron star of B $\sim$ 10$^{13}$ G
\citep{2002ApJ...580.1043B,2005Truemper}. A similar magnetic field strength of 
(2.8$-$4.2) x 10$^{13}$ G was inferred from the pulse period history of RX\,J0720.4$-$3125
as observed with ROSAT, Chandra and XMM-Newton over a time span of 10 years
\citep{2004MNRAS.351.1099C}. These were the first indications that the group of thermal 
isolated neutron stars possess strong magnetic fields of the order of $10^{13}-10^{14}$ G.
Such strong fields are indeed required to spin down the neutron stars to their current 
long rotation periods within $10^{5}-10^{6}$ years (still being sufficiently 
hot to be detected in X-rays) if they were born with msec periods.

Cyclotron resonance absorption features in the 0.1$-$1.0 keV band are expected in X-ray 
spectra from magnetized neutron stars with field strengths in the range of 10$^{10} - 10^{11}$ G
or 2 x 10$^{13} - 2$ x 10$^{14}$ G if caused by electrons or protons, respectively 
\citep[see e.g.][]{2001ApJ...560..384Z,2002nsps.conf..263Z}. 
Variation of the magnetic field strength over the neutron star surface (as expected for dipole fields)
leads to a broadening of the line \citep{2004ApJ...607..420H}. 
The strong magnetic fields inferred from magnetic dipole breaking effects in at least two 
of the stars suggests that the broad absorption features seen in the X-ray spectra of thermal 
isolated neutron stars originate from cyclotron resonance absorption by protons or highly ionized
atoms of heavy elements. With a mass to charge ratio of $\sim$2 with respect to protons the latter
case would lead to B a factor of $\sim$2 higher than that derived for protons. 
Different ionization states would result in a series of lines with energies differing 
by only a few percent, leading to additional broadening of the lines. 

An alternative possibility for the origin of the absorption line is atomic bound-bound 
transitions. In strong magnetic fields atomic orbitals are distorted into a cylindrical 
shape and the electron energy levels are similar to Landau states, with binding energies of atoms 
significantly increased. E.g. for hydrogen in a magnetic field of 
the order of 10$^{13}$ G the strongest atomic transition is expected at energy 
E/eV $\approx$ 75(1+0.13ln(B$_{13}$))+63B$_{13}$,
with B$_{13}$ = B/10$^{13}$ G \citep{2002nsps.conf..263Z}. For the line energies found 
in the spectra of thermal isolated neutron stars this would require similar field strengths
to those derived assuming cyclotron absorption.
Atomic line transitions are expected to be less prominent at higher
temperatures because of a higher ionization degree \citep{2002nsps.conf..263Z}. 

\section{Conclusions}

Although the true origin of the broad absorption lines in X-ray spectra of thermal isolated
neutron stars is not clear yet, our current knowledge about the ``magnificent seven" 
strongly suggests that they are highly magnetized ($10^{13} - 10^{14}$ G), slowly rotating
cooling neutron stars. Further timing studies would be very useful to obtain more independent
estimates of the magnetic field strength (as they currently only exist from RX\,J0720.4$-$3125).

We do not detect radio emission, probably because their radio beam is very
narrow due to their large light cylinder radius. The discovery of a few radio pulsars with similar
magnetic field strength and long period 
\citep{2000ApJ...541..367C,2002MNRAS.335..275M,2003ApJ...591L.135M} shows that radio emission can
still occur at inferred field strengths higher than the ``quantum critical field" 
$B_{cr}= m_e^2c^3/e\hbar \simeq 4.4\times 10^{13}$ G. On the other hand, any non-thermal 
emission from the ``magnificent seven" may so far just fall below the detection threshold of 
current instruments \citep{2005astroph0503239}.

\begin{acknowledgements}
 The XMM-Newton project is supported by the Bundesministerium f\"ur Bildung und
 For\-schung / Deutsches Zentrum f\"ur Luft- und Raumfahrt (BMBF / DLR), the
 Max-Planck-Gesellschaft and the Heidenhain-Stif\-tung. 
\end{acknowledgements}

\bibliographystyle{aa}


\begin{thebibliography}{36}
\expandafter\ifx\csname natexlab\endcsname\relax\def\natexlab#1{#1}\fi

\bibitem[{{Braje} \& {Romani}(2002)}]{2002ApJ...580.1043B}
{Braje}, T.~M. \& {Romani}, R.~W. 2002, \apj, 580, 1043

\bibitem[{{Burwitz} {et~al.}(2003){Burwitz}, {Haberl}, {Neuh{\"a}user},
  {Predehl}, {Tr{\"u}mper}, \& {Zavlin}}]{2003A&A...399.1109B}
{Burwitz}, V., {Haberl}, F., {Neuh{\"a}user}, R., {et~al.} 2003, \aap, 399,
  1109

\bibitem[{{Camilo} {et~al.}(2000){Camilo}, {Kaspi}, {Lyne}, {Manchester},
  {Bell}, {D'Amico}, {McKay}, \& {Crawford}}]{2000ApJ...541..367C}
{Camilo}, F., {Kaspi}, V.~M., {Lyne}, A.~G., {et~al.} 2000, \apj, 541, 367

\bibitem[{{Cropper} {et~al.}(2004){Cropper}, {Haberl}, {Zane}, \&
  {Zavlin}}]{2004MNRAS.351.1099C}
{Cropper}, M., {Haberl}, F., {Zane}, S., \& {Zavlin}, V.~E. 2004, \mnras, 351,
  1099

\bibitem[{{Cropper} {et~al.}(2001){Cropper}, {Zane}, {Ramsay}, {Haberl}, \&
  {Motch}}]{2001A&A...365L.302C}
{Cropper}, M., {Zane}, S., {Ramsay}, G., {Haberl}, F., \& {Motch}, C. 2001,
  \aap, 365, L302

\bibitem[{{de Vries} {et~al.}(2004){de Vries}, {Vink}, {M{\'e}ndez}, \&
  {Verbunt}}]{2004A&A...415L..31D}
{de Vries}, C.~P., {Vink}, J., {M{\'e}ndez}, M., \& {Verbunt}, F. 2004, \aap,
  415, L31

\bibitem[{{G{\" a}nsicke} {et~al.}(2002){G{\" a}nsicke}, {Braje}, \&
  {Romani}}]{2002A&A...386.1001G}
{G{\" a}nsicke}, B.~T., {Braje}, T.~M., \& {Romani}, R.~W. 2002, \aap, 386,
  1001

\bibitem[{{Haberl}(2004)}]{haberl2004COSPAR}
{Haberl}, F. 2004, Adv. Space Res., 33, 638

\bibitem[{{Haberl} {et~al.}(1997){Haberl}, {Motch}, {Buckley}, {Zickgraf}, \&
  {Pietsch}}]{1997A&A...326..662H}
{Haberl}, F., {Motch}, C., {Buckley}, D. A.~H., {Zickgraf}, F.~J., \&
  {Pietsch}, W. 1997, \aap, 326, 662

\bibitem[{{Haberl} {et~al.}(2004{\natexlab{a}}){Haberl}, {Motch}, {Zavlin},
  {Reinsch}, {G{\" a}nsicke}, {Cropper}, {Schwope}, {Turolla}, \&
  {Zane}}]{2004A&A...424..635H}
{Haberl}, F., {Motch}, C., {Zavlin}, V.~E., {et~al.} 2004{\natexlab{a}}, \aap,
  424, 635

\bibitem[{{Haberl} {et~al.}(2003){Haberl}, {Schwope}, {Hambaryan}, {Hasinger},
  \& {Motch}}]{2003A&A...403L..19H}
{Haberl}, F., {Schwope}, A.~D., {Hambaryan}, V., {Hasinger}, G., \& {Motch}, C.
  2003, \aap, 403, L19

\bibitem[{{Haberl} \& {Zavlin}(2002)}]{2002A&A...391..571H}
{Haberl}, F. \& {Zavlin}, V.~E. 2002, \aap, 391, 571

\bibitem[{{Haberl} {et~al.}(2004{\natexlab{b}}){Haberl}, {Zavlin},
  {Tr{\"u}mper}, \& {Burwitz}}]{2004A&A...419.1077H}
{Haberl}, F., {Zavlin}, V.~E., {Tr{\"u}mper}, J., \& {Burwitz}, V.
  2004{\natexlab{b}}, \aap, 419, 1077

\bibitem[{{Hambaryan} {et~al.}(2002){Hambaryan}, {Hasinger}, {Schwope}, \&
  {Schulz}}]{2002A&A...381...98H}
{Hambaryan}, V., {Hasinger}, G., {Schwope}, A.~D., \& {Schulz}, N.~S. 2002,
  \aap, 381, 98

\bibitem[{{Ho} \& {Lai}(2004)}]{2004ApJ...607..420H}
{Ho}, W.~C.~G. \& {Lai}, D. 2004, \apj, 607, 420

\bibitem[{{Kaplan} {et~al.}(2002){Kaplan}, {Kulkarni}, \& {van
  Kerkwijk}}]{2002ApJ...579L..29K}
{Kaplan}, D.~L., {Kulkarni}, S.~R., \& {van Kerkwijk}, M.~H. 2002, \apjl, 579,
  L29

\bibitem[{{Kaplan} {et~al.}(2003){Kaplan}, {Kulkarni}, \& {van
  Kerkwijk}}]{2003ApJ...588L..33K}
{Kaplan}, D.~L., {Kulkarni}, S.~R., \& {van Kerkwijk}, M.~H. 2003, \apjl, 588,
  L33

\bibitem[{{McLaughlin} {et~al.}(2003){McLaughlin}, {Stairs}, {Kaspi},
  {Lorimer}, {Kramer}, {Lyne}, {Manchester}, {Camilo}, {Hobbs}, {Possenti},
  {D'Amico}, \& {Faulkner}}]{2003ApJ...591L.135M}
{McLaughlin}, M.~A., {Stairs}, I.~H., {Kaspi}, V.~M., {et~al.} 2003, \apjl,
  591, L135

\bibitem[{{Morris} {et~al.}(2002){Morris}, {Hobbs}, {Lyne}, {Stairs}, {Camilo},
  {Manchester}, {Possenti}, {Bell}, {Kaspi}, {Amico}, {McKay}, {Crawford}, \&
  {Kramer}}]{2002MNRAS.335..275M}
{Morris}, D.~J., {Hobbs}, G., {Lyne}, A.~G., {et~al.} 2002, \mnras, 335, 275

\bibitem[{{Motch}(2001)}]{2001xase.conf..244M}
{Motch}, C. 2001, in X-ray Astronomy, Stellar Endpoints, AGN, and the Diffuse
  X-ray Background, AIP Conference Proceedings, 244--253

\bibitem[{{Motch} {et~al.}(1999){Motch}, {Haberl}, {Zickgraf}, {Hasinger}, \&
  {Schwope}}]{1999A&A...351..177M}
{Motch}, C., {Haberl}, F., {Zickgraf}, F.~J., {Hasinger}, G., \& {Schwope},
  A.~D. 1999, \aap, 351, 177

\bibitem[{{Motch} {et~al.}(2005){Motch}, {Sekiguchi}, {Haberl}, {Zavlin},
  {Schwope}, \& {Pakull}}]{2005A&A...429..257M}
{Motch}, C., {Sekiguchi}, K., {Haberl}, F., {et~al.} 2005, \aap, 429, 257

\bibitem[{{Motch} {et~al.}(2003){Motch}, {Zavlin}, \&
  {Haberl}}]{2003A&A...408..323M}
{Motch}, C., {Zavlin}, V.~E., \& {Haberl}, F. 2003, \aap, 408, 323

\bibitem[{{Pavlov} {et~al.}(2002){Pavlov}, {Zavlin}, \&
  {Sanwal}}]{2002nsps.conf..273P}
{Pavlov}, G.~G., {Zavlin}, V.~E., \& {Sanwal}, D. 2002, in Neutron Stars,
  Pulsars, and Supernova Remnants, Eds. W. Becker, H. Lesch and J. Tr\"umper,
  MPE-Report 278, 273--286

\bibitem[{{Schwope} {et~al.}(2005){Schwope}, {Hambaryan}, {Haberl}, \&
  {Motch}}]{2005A&A...INS}
{Schwope}, A.~D., {Hambaryan}, V., {Haberl}, F., \& {Motch}, C. 2005, \aap\ in
  press

\bibitem[{{Schwope} {et~al.}(1999){Schwope}, {Hasinger}, {Schwarz}, {Haberl},
  \& {Schmidt}}]{1999A&A...341L..51S}
{Schwope}, A.~D., {Hasinger}, G., {Schwarz}, R., {Haberl}, F., \& {Schmidt}, M.
  1999, \aap, 341, L51

\bibitem[{{Treves} {et~al.}(2000){Treves}, {Turolla}, {Zane}, \&
  {Colpi}}]{2000PASP..112..297T}
{Treves}, A., {Turolla}, R., {Zane}, S., \& {Colpi}, M. 2000, \pasp, 112, 297

\bibitem[{{Tr{\"u}mper}(2005)}]{2005Truemper}
{Tr{\"u}mper}, J.~E. 2005, in to appear in ASI proceedings of The
  Electromagnetic Spectrum of Neutron Stars, Marmaris 2004 {(astro-ph/0502457)}

\bibitem[{{van Kerkwijk} {et~al.}(2004){van Kerkwijk}, {Kaplan}, {Durant},
  {Kulkarni}, \& {Paerels}}]{2004ApJ...608..432V}
{van Kerkwijk}, M.~H., {Kaplan}, D.~L., {Durant}, M., {Kulkarni}, S.~R., \&
  {Paerels}, F. 2004, \apj, 608, 432

\bibitem[{{van Kerkwijk} \& {Kulkarni}(2001)}]{2001A&A...380..221V}
{van Kerkwijk}, M.~H. \& {Kulkarni}, S.~R. 2001, \aap, 380, 221

\bibitem[{{Walter} \& {Lattimer}(2002)}]{2002ApJ...576L.145W}
{Walter}, F.~M. \& {Lattimer}, J.~M. 2002, \apjl, 576, L145

\bibitem[{{Walter} \& {Matthews}(1997)}]{1997Natur.389..358W}
{Walter}, F.~M. \& {Matthews}, L.~D. 1997, \nat, 389, 358

\bibitem[{{Zampieri} {et~al.}(2001){Zampieri}, {Campana}, {Turolla},
  {Chieregato}, {Falomo}, {Fugazza}, {Moretti}, \&
  {Treves}}]{2001A&A...378L...5Z}
{Zampieri}, L., {Campana}, S., {Turolla}, R., {et~al.} 2001, \aap, 378, L5

\bibitem[{{Zane} {et~al.}(2005){Zane}, {Cropper}, {Turolla}, {Zampieri},
  {Chieregato}, {Drake}, \& {Treves}}]{2005astroph0503239}
{Zane}, S., {Cropper}, M., {Turolla}, R., {et~al.} 2005, \apj, in press
  {(astro-ph/0503239)}

\bibitem[{{Zane} {et~al.}(2001){Zane}, {Turolla}, {Stella}, \&
  {Treves}}]{2001ApJ...560..384Z}
{Zane}, S., {Turolla}, R., {Stella}, L., \& {Treves}, A. 2001, \apj, 560, 384

\bibitem[{{Zavlin} \& {Pavlov}(2002)}]{2002nsps.conf..263Z}
{Zavlin}, V.~E. \& {Pavlov}, G.~G. 2002, in Neutron Stars, Pulsars, and
  Supernova Remnants, Eds. W. Becker, H. Lesch and J. Tr\"umper, MPE-Report
  278, 263--272

\end{thebibliography}

\end{document}